\documentclass[english,twocolumn,superscriptaddress,longbibliography,floatfix]{revtex4-1}

\usepackage{graphicx}
\usepackage{amsmath,amsfonts}
\usepackage{hyperref}
\usepackage{tikz}
\usepackage{siunitx}
\usepackage{float}
\usepackage[title]{appendix}
\usepackage{xcolor}

\newcommand{\bs}[1]{\boldsymbol{#1}}

\definecolor{lime}{HTML}{A6CE39}
\DeclareRobustCommand{\orcidicon}{
  \begin{tikzpicture}
  \draw[lime, fill=lime] (0,0)
  circle [radius=0.16]
  node[white] {{\fontfamily{qag}\selectfont \tiny ID}};
  \draw[white, fill=white] (-0.0625,0.095)
  circle [radius=0.007];
  \end{tikzpicture}
  \hspace{-2mm}
}
\foreach \x in {A, ..., Z}{\expandafter\xdef\csname orcid\x\endcsname{\noexpand\href{https://orcid.org/\csname
      orcidauthor\x\endcsname}{\noexpand\orcidicon}} }

\begin{document}

\title{Online optimization of nonlinear lattice using a data-driven chaos indicator}

\author{Minghao Song\orcidA{}}
\author{Yongjun Li\orcidB{}}\email{yli@bnl.gov}
\affiliation{Brookhaven National Laboratory, Upton, New York 11973, USA}

\begin{abstract}
We report the experimental implementation of a Data-Driven Chaos Indicator (DDCI) [Y.~Li \emph{et al.}, Nucl.\ Instrum.\ Methods Phys.\ Res.\ A \textbf{1024} (2022) 166060] for online optimization of the National Synchrotron Light Source II (NSLS-II) storage ring. The DDCI quantifies the predictability of electron beam dynamics using turn-by-turn beam position monitor data. A surrogate model of the one-turn map is first trained, and its out-of-sample predictive uncertainty is then employed as a measurable indicator of chaos. By tuning sextupole magnets to mitigate nonlinear effects, a clear enlargement of the dynamic aperture is achieved, accompanied by a corresponding improvement in injection efficiency.
\end{abstract}

\maketitle

The nonlinear beam dynamics of storage rings determine key operational performance metrics, including injection efficiency and beam lifetime, which are closely correlated with the dynamic aperture (DA) and momentum acceptance (MA). During the design stage, storage ring lattices are typically optimized to enhance DA and MA using physics-based accelerator models and/or simulation-driven objectives. However, the ``as-built" machine inevitably contains unmodeled imperfections, such as magnet field errors, alignment and calibration residuals, hysteresis effects, and stray fields. These discrepancies can significantly degrade nonlinear performance relative to design expectations. Consequently, online optimization based directly on experimental data is strongly motivated.

A key challenge in the online optimization of nonlinear lattices lies in defining a robust objective function that (1) is experimentally measurable within limited beam time, (2) is sensitive to the nonlinear dynamics governing DA and MA, and (3) is sufficiently smooth and repeatable to effectively guide an optimizer under operational constraints. Traditional chaos indicators, such as Lyapunov exponents~\cite{LE_simplectic,Schmidt_LE,fischer1995experimental}, forward--reverse integration error~\cite{Li_forward_backward,Federico_REM}, and tangent map methods~\cite{Skokos_SALI_2001,Qiang_tangentmap}, are powerful in simulation but are not readily quantifiable in real-world accelerators.

This paper presents the experimental validation of a Data-Driven Chaos Indicator (DDCI), originally introduced and demonstrated offline using simulated data in Ref.~\cite{li2022data}. The method exploits the fundamental property that chaos limits predictability. When beam motion is represented by turn-by-turn trajectories, predictability can be quantified by training a surrogate model to learn the one-turn map from beam position monitor (BPM) data and evaluating its out-of-sample prediction performance. For a fixed surrogate architecture and data quality, reduced predictability reflects increased chaos in the underlying system. This behavior can be leveraged as an objective for optimization by tuning the available control knobs.

Consider the transverse one-turn map of the storage ring,
\begin{equation*}
\bs{X}_{n+1} = M(\bs{X}_n), 
\qquad 
\bs{X}=(x,x',y,y'),
\label{eq:oneturnmap}
\end{equation*}
where $M$ denotes the effective one-turn transformation at the observation point and $\bs{X}_n$ represents the transverse phase-space coordinate at turn $n$ with the closed-orbit components subtracted. In an ideal symplectic description, the map $M$ is deterministic and measure-preserving for single-particle motion. Chaos is then characterized by exponential sensitivity to initial conditions, meaning that small differences in initial conditions can grow exponentially over time. In the linear regime, two close initial conditions, separated by $\Delta \bs{X}_n$, propagate linearly:
\begin{equation}
\Delta \bs{X}_{n+1} 
\approx 
J_M(\bs{X}_n)\,\Delta \bs{X}_n,\nonumber
\end{equation}
where $J_M$ is the Jacobian of $M$. However, for chaotic trajectories, the deviation evolves approximately as
\begin{equation}
\|\Delta \bs{X}_{n+1}\| 
\sim 
e^{\lambda \tau}\,\|\Delta \bs{X}_n\|,\nonumber
\end{equation}
with $\lambda$ denoting a finite-time Lyapunov exponent (with $\tau$ the revolution time of electrons in storage rings). Even without explicitly computing $\lambda$, this exponential growth implies a limited prediction window beyond which accurate prediction of the trajectory becomes unreliable.

Under real machine conditions, the observed map also includes measurement and dynamical effects. When we attempt to predict $\bs{X}_{n+1}$ from $\bs{X}_n$ using a surrogate $\widehat{M}$ trained from data, the one-turn prediction error can be approximately decomposed into three contributions:
\begin{equation}
\begin{aligned}
\bs{e}_{n+1} &\equiv \bs{X}_{n+1}-\widehat{M}(\bs{X}_n) \\
&= \underbrace{\left[M(\bs{X}_n)-\widehat{M}(\bs{X}_n)\right]}_{\text{model bias}}
+ \underbrace{\bs{r}_{n+1}}_{\substack{\text{random} \\ \text{effects}}}
+ \underbrace{\bs{d}_{n+1}}_{\substack{\text{slow} \\ \text{drifts}}},
\end{aligned}
\label{eq:error-decomp}
\end{equation}
where $\bs{r}_{n+1}$ represents measurement noise, for example, fluctuations in BPM readings, radiation damping, and decoherence. $\bs{d}_{n+1}$ describes slow drifts, such as changes in orbit or optics over the evaluation window. Under the present experimental conditions, measurement noise and slow drift are mitigated by high-precision beam diagnostics and a confined acquisition window, while the surrogate structure is fixed across evaluations. For optimization, variations in prediction error are therefore dominated by the underlying nonlinear dynamics, allowing predictability to serve as a practical proxy for chaos.

When the oscillation amplitude is sufficiently small, nearby points in phase space evolve smoothly, so the map $M$ can be well approximated by a surrogate of fixed complexity, leading to a small bias. However, as the amplitude increases, the map exhibits stronger chaos, and a progressively more complex surrogate is needed. If we insist on using a fixed-complexity surrogate, chaotic motion can be characterized as irreducible error.

The use of predictability as a practical chaos indicator can be further justified by the exponential amplification of uncertainty and finite data resolution. In chaotic systems, small uncertainties grow exponentially over time, quickly limiting predictive accuracy. At the same time, finite data make learning more difficult, especially as nearby points in phase space diverge and follow different dynamics. As a result, reduced predictability reflects stronger chaotic behavior. Under fixed surrogate class, data acquisition, and noise conditions, variations in prediction error primarily reflect changes in the underlying dynamical complexity rather than modeling artifacts.

Although predictability is not the same as a Lyapunov exponent, in our case, the DDCI serves as a consistent measure of how difficult the underlying dynamics are to learn and predict. In principle, long-term predictability quantifies chaos accurately. In practice, in storage rings with strong lattice structure, local nonlinear distortions captured in the one-turn map are strongly correlated with global stability properties such as DA. Therefore, one-turn predictability is already informative for online tuning at real-world accelerators.

The above considerations establish predictability as a practical and measurable chaos indicator. We now describe its experimental implementation. The experiment is conducted on the NSLS-II storage ring, a 3\,GeV electron storage ring comprising 30 cells with a double-bend achromat lattice~\cite{dierker2007}. While its baseline nonlinear lattice was optimized during the design stage, the operational machine exhibits residual nonlinear performance limitations. In this study, six families of harmonic sextupoles are used as tuning knobs to suppress chaos without altering the linear chromaticity.

For each sextupole configuration evaluated by the optimizer, the beam is excited with a pinger magnet to four progressively increasing amplitudes in the transverse $(x\text{--}y)$ plane to probe nonlinearities. At each amplitude, three repeated excitations are performed to assess repeatability and provide multiple datasets for training and testing statistics. Then a dedicated BPM pair separated by a pure drift is used to acquire turn-by-turn (TbT) beam positions~\cite{Li_dedicated_BPMs}, which are then used to construct the four-dimensional phase-space coordinate after nonlinear calibration~\cite{cheng2012nsls2}. Instead of generating many independent trajectories, successive turns from a single excitation are used to form input--output pairs. The main reason is that the measurement can be completed within a finite operational time window. In the meantime, the slow drift in Eq.~\eqref{eq:error-decomp} can be neglected.

\begin{figure}[t]
  \centering
  \includegraphics[width=0.8\columnwidth]{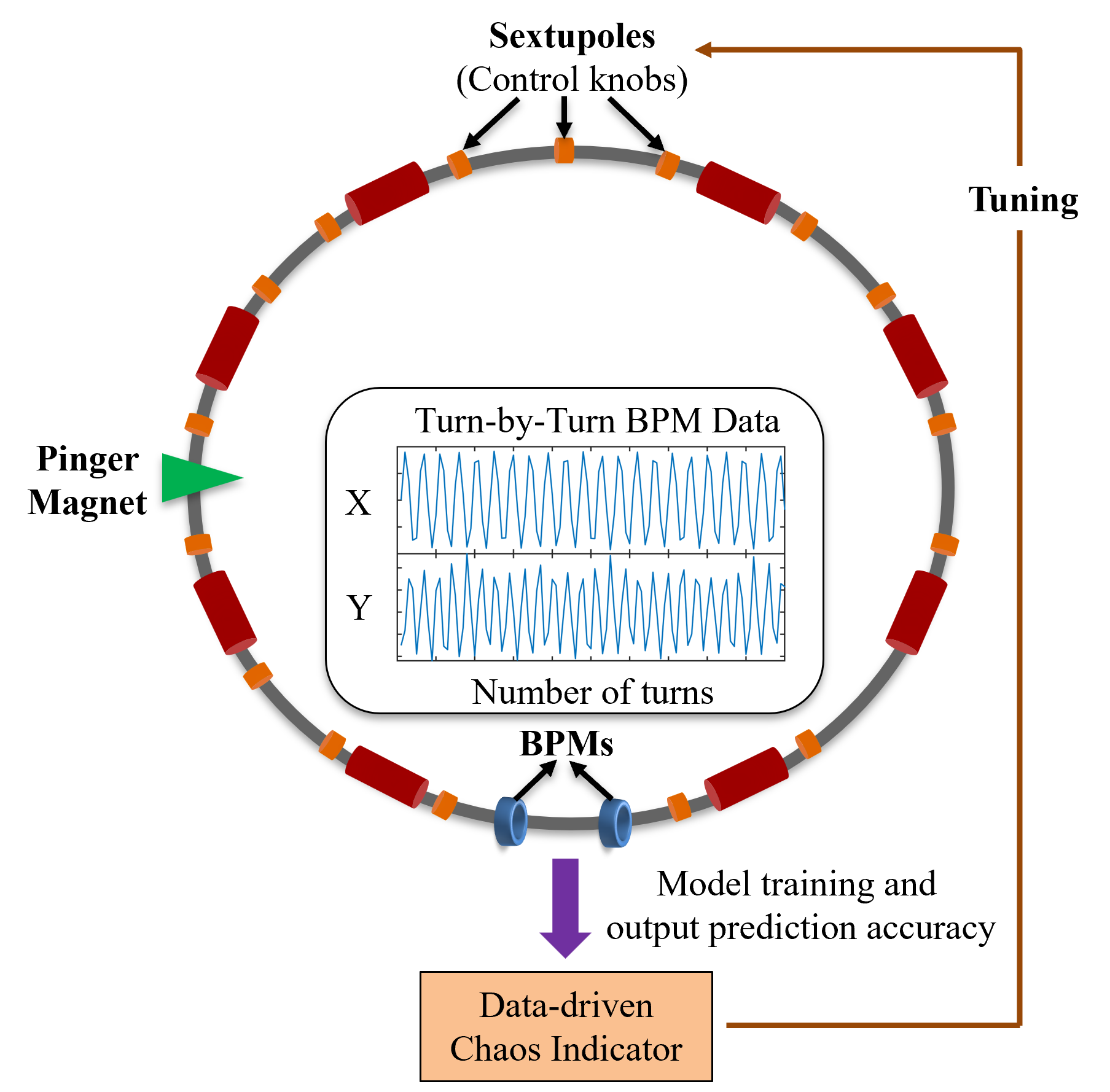}
  \caption{Schematic of the online optimization loop using DDCI. For each sextupole setting, the stored beam is excited by the pinger magnet and turn-by-turn data are recorded by a BPM pair. The reconstructed phase-space samples form one-turn input--output pairs to train a fixed-structure surrogate. The surrogate's out-of-sample predictive accuracy defines the DDCI objective. Bayesian optimization proposes new sextupole settings to minimize the DDCI, and the loop repeats until convergence.}
  \label{fig:optimization_loop}
\end{figure}

The surrogate is obtained by training on a randomly selected ~90\% subset of input--output pairs, while the remaining pairs are used for out-of-sample prediction testing. The prediction performance is quantified as a chaos indicator to drive the optimizer to adjust the sextupole knobs. To assess robustness against sampling variance, each dataset is randomly shuffled for cross-validation. The complete closed-loop optimization procedure is illustrated in Fig.~\ref{fig:optimization_loop}.

\begin{figure}[t]
  \centering
  \includegraphics[width=0.95\columnwidth]{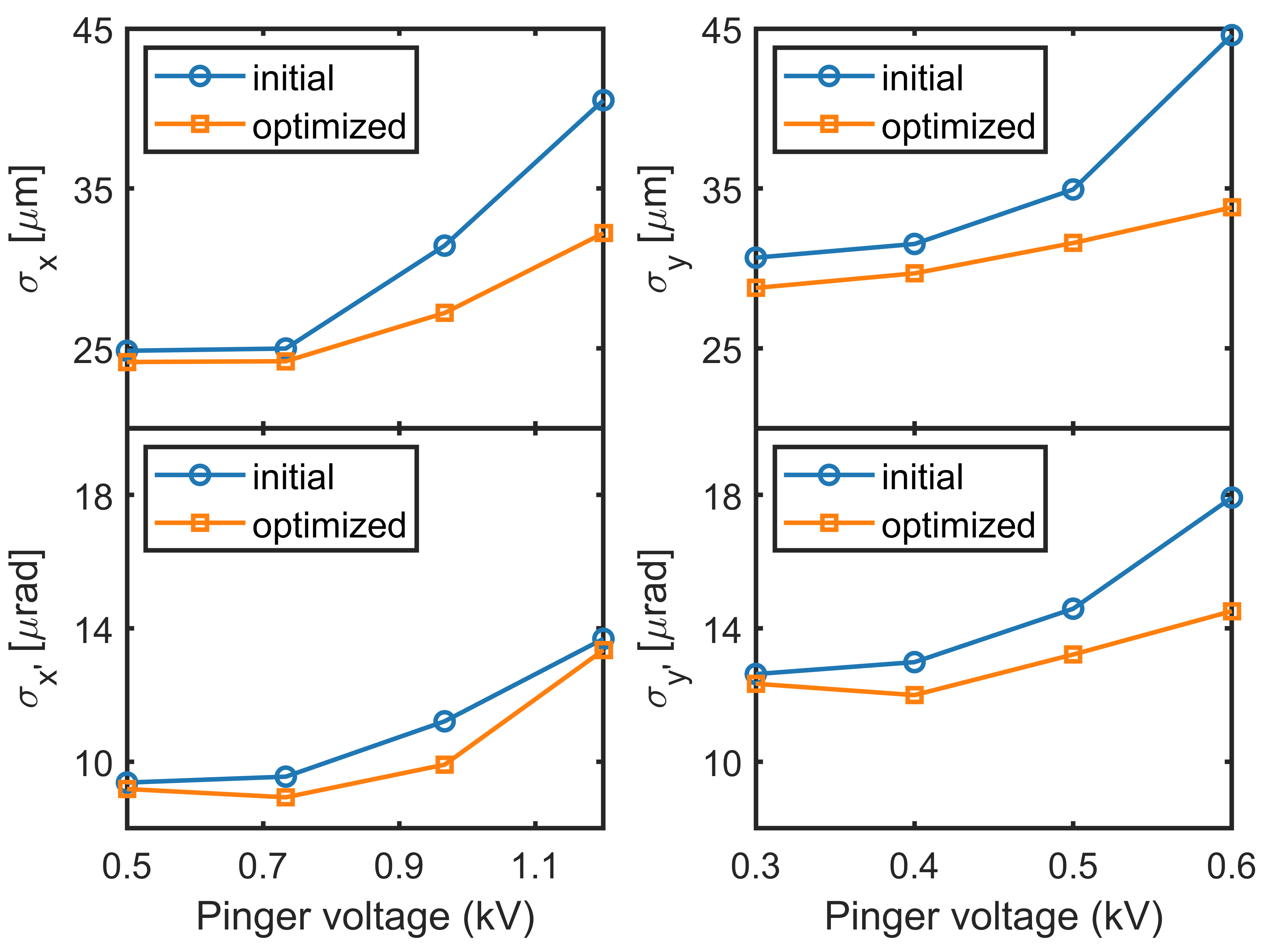}
  \caption{Mean standard deviation of the prediction errors for $(x,x',y,y')$ at four excitation amplitudes (pinger voltages) for the initial sextupole configuration and the optimized one. Prediction errors are computed on out-of-sample test sets with a 90\%/10\% split. The optimized setting reduces the DDCI across amplitudes, with relatively larger reduction at higher excitation where nonlinear effects are stronger.}
  \label{fig:cmp_uncertainty_vs_amp}
\end{figure}

Instead of using a neural network as the surrogate in Ref.~\cite{li2022data}, we adopt a Gaussian process regression model~\cite{williams2006gaussian} to demonstrate the model-independence of the proposed approach. Accordingly, the specific choice of surrogate $\widehat{M}$ is not critical, provided that its structural complexity is fixed for evaluating predictive performance. The DDCI objective, for a given excitation amplitude, is quantified by the standard deviation of the prediction errors on an out-of-sample test set,
\begin{equation*}
 \sigma_{q} = \sqrt{\frac{1}{N}\sum_{j=1}^N\Delta q_{j}^2},\quad \Delta q = q_{\mathrm{meas.}} - q_{\mathrm{pred.}},
\end{equation*}
where $N$ denotes the total number of test set, and $q\in\bs{X}$. To suppress chaos across amplitudes and mitigate scale differences, for each sextupole configuration, its DDCI is normalized by the pre-selected baseline values $\sigma_{q,0}$, which are measured with the initial sextupole settings. The total objective minimized by the optimizer is
\begin{equation*}
\sigma  = \sum_{i=1}^4\sum_{q\in \bs{X}} 
\left(\frac{\sigma_{q}} {\sigma_{q,0}}\right)_i, 
\label{eq:objective}
\end{equation*}
where, $i=1,\ldots,4$ denotes four progressively increasing pinger excitation amplitudes, uniformly spanning the ranges $[0.5,1.2]\,\mathrm{kV}$ in the horizontal plane and $[0.3,0.6]\,\mathrm{kV}$ in the vertical plane, as illustrated in Fig.~\ref{fig:cmp_uncertainty_vs_amp}. Lower $\sigma$ indicates higher predictability, i.e., more regular beam motion. Equal weighting is adopted to avoid bias toward any single phase-space coordinate and to provide a balanced measure of overall predictability.

A Bayesian optimizer was adopted due to its suitability for optimizing noisy, expensive-to-evaluate objective functions under a limited evaluation budget. Each optimization epoch typically required approximately $40$ evaluations, corresponding to about one hour of machine time. This duration aligns well with routine user-facility operations, during which slow system drift is also negligible. After optimization, the prediction performance is  improved across the four pinger excitations, as shown in Fig.~\ref{fig:cmp_uncertainty_vs_amp}, which is consistent with stronger sensitivity to nonlinearities at larger oscillation amplitudes.

\begin{figure}[t]
  \centering
  \includegraphics[width=0.95\columnwidth]{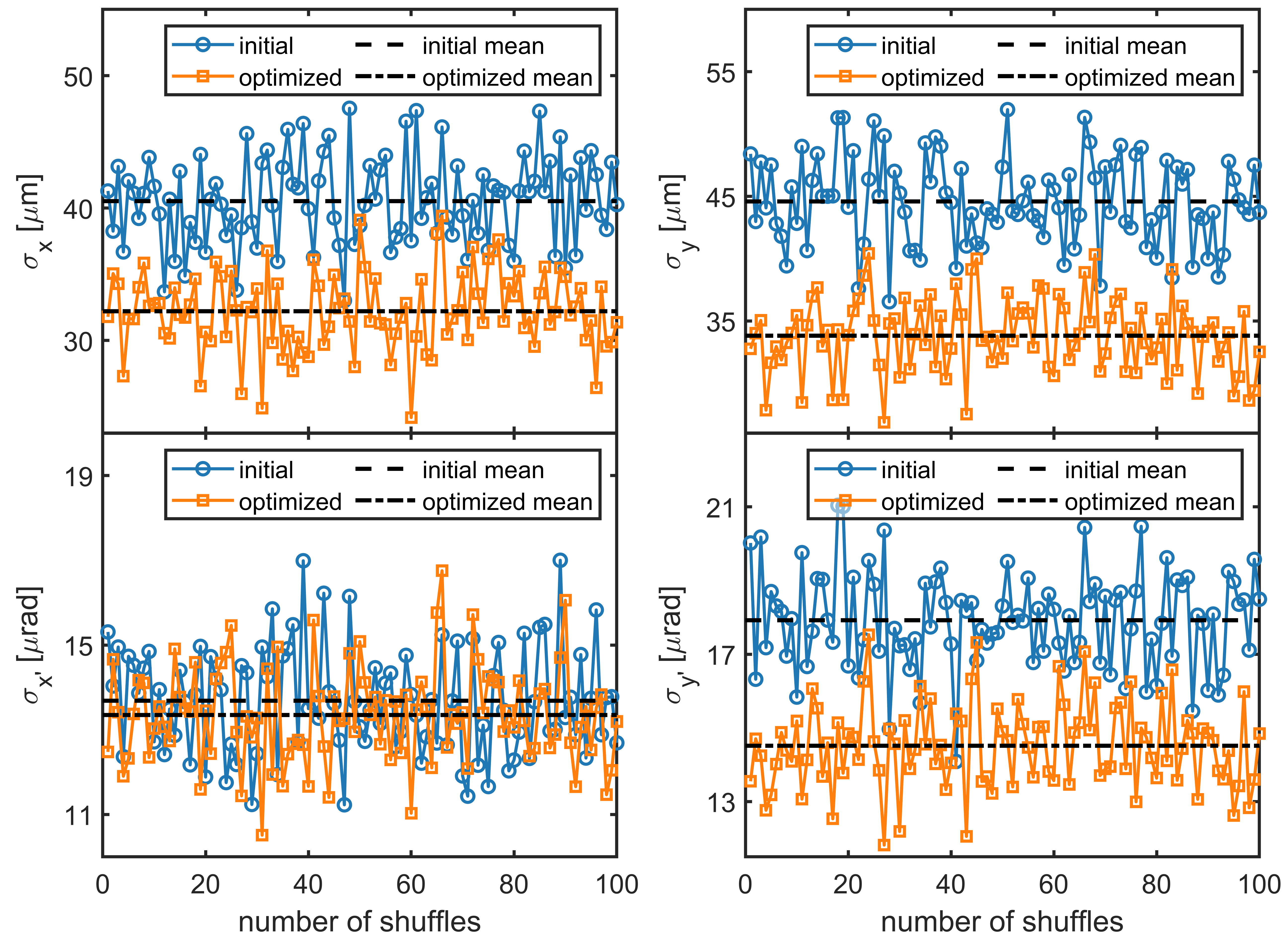}
  \caption{Cross validation with 100 random shuffles for the largest excitation amplitude (1.2\,kV horizontal, 0.6\,kV vertical). The optimized configuration consistently exhibits reduced standard deviation of the prediction errors across phase-space coordinates. Variations among individual coordinates reflect differences in sensitivity to nonlinear dynamics and measurement noise.}
  \label{fig:cmp_shuffle}
\end{figure}

To confirm that the observed reduction in the DDCI is not an artifact of a specific data partition, but instead reflects systematic chaos suppression, cross-validation, as shown in Fig.~\ref{fig:cmp_shuffle}, was adopted. While most phase-space coordinates exhibit a clear reduction in predictive uncertainty, one component $x'$ shows weaker improvement and larger statistical fluctuations. This is acceptable, as the optimization minimizes an aggregate objective rather than individual coordinates, and different phase-space directions can have varying sensitivity to sextupole tuning and measurement noise. Despite this, the overall reduction in the aggregated DDCI and the independently observed improvement in DA, as demonstrated in the following measurements, confirm the effectiveness of the optimization.

\begin{figure}[t]
  \centering
  \includegraphics[width=0.8\columnwidth]{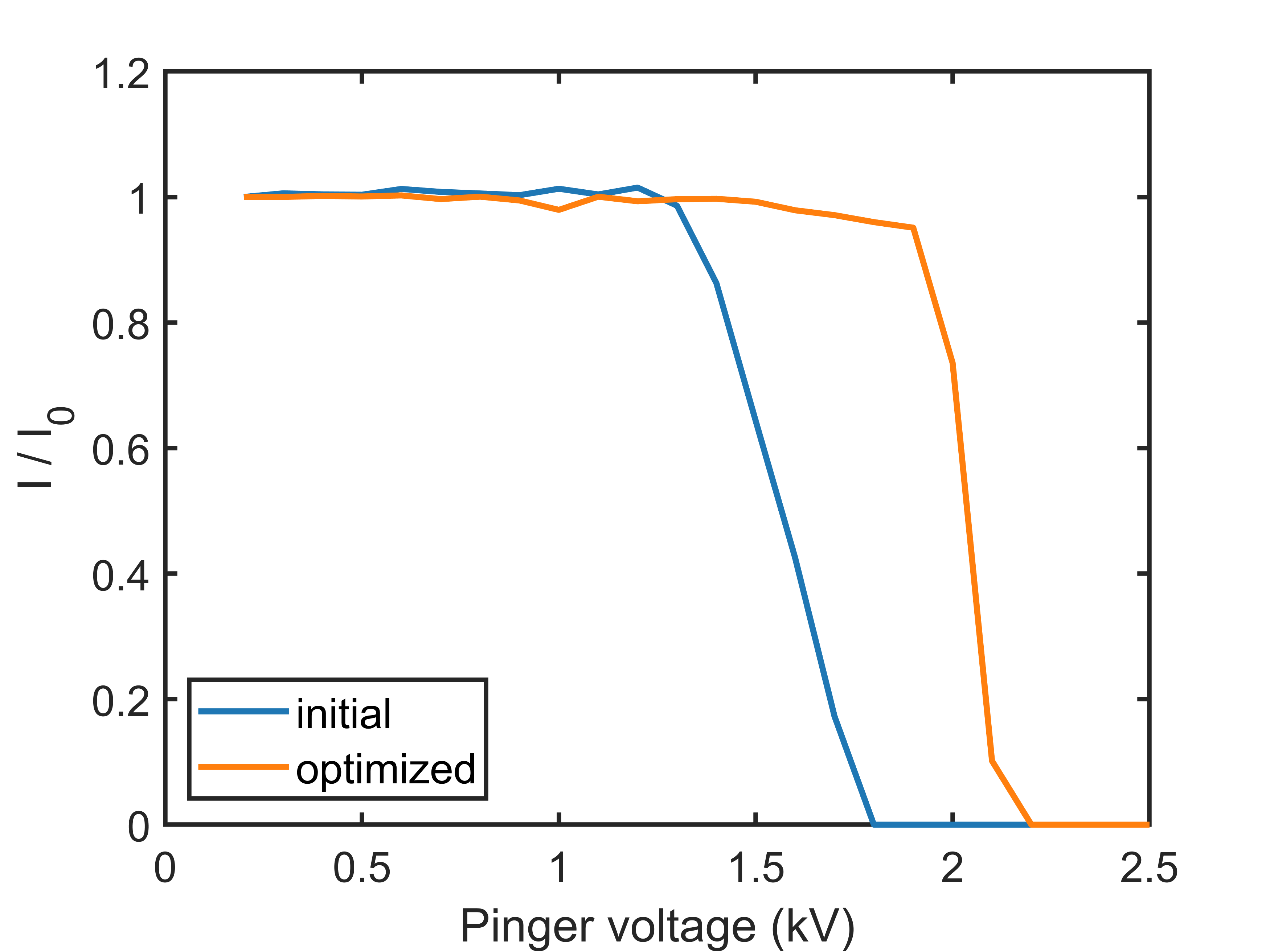}
  \caption{Measurement of the horizontal DA with the initial/optimized sextupole configurations. The stored beam is kicked with progressively increasing pinger voltage until beam loss occurs. The optimized configuration can tolerate larger kicks, indicating an enlarged DA.}
  \label{fig:cmp_DA}
\end{figure}

With the optimized configuration, an enlarged DA is confirmed by a direct measurement. In this measurement, the pinger kick strength is progressively increased until the stored beam is completely lost from the DA. Figure~\ref{fig:cmp_DA} compares the beam-loss patterns for the initial and optimized configurations. Under the same measurement protocol, the optimized setting sustains larger kick amplitudes before beam loss occurs, demonstrating a significant enlargement of the horizontal DA. In line with the improvement in DA, the off-axis injection efficiency--formerly adopted as the optimization objective~\cite{huang2019beam}--improves from about $70\%$ to above $90\%$ after optimization. The beam lifetime remains similar under identical bunch filling patterns and beam current conditions, suggesting that the optimization improves transverse DA without degrading MA.

The online optimization was repeated under multiple operational configurations, including with all insertion device gaps open and engaged. In all cases, improvements in DA and injection efficiency were consistently achieved, demonstrating the robustness and operational reliability of this accelerator-model-independent online approach.

We note the differences between using DDCI in simulation~\cite{li2022data} and measurement. In offline optimization with simulation data, the one-turn map is obtained from a high-fidelity accelerator model, and a large ensemble of independent trajectories can be generated in parallel across phase space. The surrogate is trained on abundant synthetic data with controlled initial conditions, and chaos arises purely from the dynamical system itself. In contrast, online optimization uses measurement data that include random errors, a limited data pool, and a finite time window to complete each optimization loop. Therefore, although the theoretical foundation of DDCI is shared between offline and online contexts, the data acquisition strategy, and robustness analysis differ substantially. 

In summary, we demonstrated online optimization of nonlinear lattice performance at NSLS-II using a DDCI derived from the predictability of turn-by-turn beam motion. Unlike the previously proposed Shannon entropy indicator~\cite{Li_Shannon_entropy}, which requires phase-space transformations, DDCI relies directly on turn-by-turn data. The timescale of each optimization epoch is compatible with operational constraints. While demonstrated on NSLS-II, the method is expected to be transferable to other facilities with turn-by-turn diagnostics owing to its model independence. Further validation on other facilities would be valuable.

\section*{Data availability}
The data that support the findings of this study  are available upon reasonable request and subject to standard U.S. national laboratory data-sharing policies.

\begin{acknowledgments}
This work is supported by the U.S. Department of Energy, Office of Basic Energy Sciences, under Contract No.~DE-SC0012704 and Field Work Proposal No.~2025-BNL-PS040.
\end{acknowledgments}

\bibliography{ref.bib}

\end{document}